# Fano Resonance Induced Anomalous Collective Hotspots in Metallic Nanoparticle Dimer Chains

**Gang Song**[1] **and Wei Zhang**[1,2,*]

[1]Institute of Applied Physics and Computation Mathematics, P. O. Box 8009(28), Beijing 100088, China

[2]Beijing Computational Science Research Centre, Beijing 100084, China

E-mail:zhang_wei@iapcm.ac.cn

**Abstract.** Hotspots with strong near fields due to localized surface plasmons (LSPs) in metallic nanostructures have various applications, such as surface enhanced Raman scattering (SERS). The long range Coulomb coupling between LSPs in periodic metallic nanostructures may lead to interesting collective effects. In this paper, we investigate the combination effects of the local field enhancement and collective plasmon resonances in one dimensional metallic nanoparticle dimer chains. It is found that the strong near field in the gap and the far field interactions among the metallic nanoparticles lead to anomalous collective hotspots with dual enhancement of the electromagnetic field. In particular, the interference between the incident field and the induced internal field leads to *Fano-type effect with Wood anomaly related destructive interference* and the strong resonance with an extremely narrow width. Our systematic study shows that the correlation between the local structure and the global structure has important impact on the collective spots, which leads to an optimal orientation of the dimer (about 60 ° with respect to the chain direction) for the largest gap field enhancement with the incident field polarization parallel to the long axis of the dimer.

## 1. Introduction

Localized surface plasmons (LSPs) on metallic nanoparticles may lead to enhancement of the near field. Strong local fields and light-matter interactions could be generated in some small regimes in composite metallic nanostructures, i.e., the hotspots. Hotspots based on surface plasmons (SPs) are widely used in many areas, such as the surface enhanced Raman scattering (SERS), and biosensors [1-8]. Several methods could be applied to generate hotspots [9-17]. For example, a bowtie or a dimer composed of two metallic nanoparticles with a small inter-face distance was used to obtain large electromagnetic fields in the gap with the polarization of the incident field along the dimer. The shape, the size of the nanoparticles, and the inter-face distance influence the hotspots.

The formation of hotspots in the bowtie or the dimer structures is due to the near field interaction. Far field interactions among nanoparticles in metallic nanoparticle arrays lead to a collective plasmon resonance [18-23]. Much effort has been devoted to the studies on the dispersion relation, transmission/reflection spectra of particle/hole arrays or grating structures [24-28]. Park and Stroud demonstrated theoretically the dispersion relation in one-dimensional (1D) nanoparticle chains [29]. The metallic nanoparticle arrays/chains could be used as waveguides for light propagating at nanoscale [30-32]. The bended arrays/chains are applied in light focusing [33, 34]. Hicks et al studied Rayleigh scattering in 1D Ag nanoparticle chains [35].

It is interesting and important to explore the structures with both near field and far field interactions involved. Hotspot in close-packed nanoparticle array was investigated in [36]. In the past decades, there were several studies on the nanoparticle dimer arrays. Field enhancement in 1D and two-dimensional dimer arrays was investigated in [37-39]. The electromagnetic field enhancement from silver nanoshell dimer arrays and the applications in SERS were discussed in [40, 41]. Some other studies also considered arrays of dimers with different dimer structure, such as slanted gap and bowtie nanostructures/nano-antenna [42-47]. Though there has been some progress in the study of nanoparticle dimer arrays, more studies are much needed to further explore the interplay between local field and collective behavior, in particular, the effect of the correlation between the local structure and the global periodic structure. In this paper, we investigate the optical properties of 1D metallic nanoparticle dimer chain of different dimer orientations and incident field polarizations, paying particular attention to the interference/correlation effect due to the interplay between the near field and the far field. The generalized coupled dipole method is developed, which incorporates both the near field and the far field interaction and works well for the case of smaller inter-face distance (1/5 the radius). It is found that the near field interaction within the dimer and the far field interaction among the dimers lead to not only the anomalous absorption, but also the collective hotspots. The combination effects of the near field concentration of light and the collective resonances result in an extremely narrow resonance width and much higher local fields in the hotspots than that of a single dimer. The relation among the interference, Fano resonance and Wood anomaly are also addressed. The tunable collective hotspots due to the correlation between the local dimer structure and the global periodic structure have various applications, such as SERS.

## 2. Calculation models and results

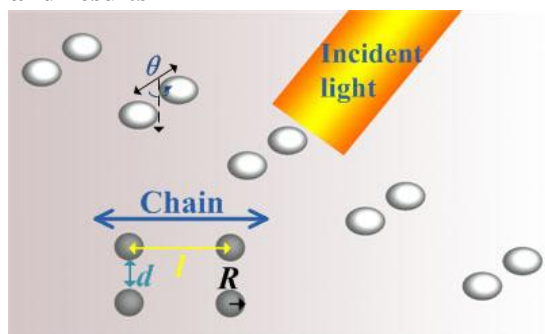

Figure 1. (Color online) The schematic diagram of the calculation models.

We systematically study the 1D silver nanoparticle dimer chains arranged in different configurations. As shown in Figure 1, the basic unit of the 1D chain is an Ag dimer with the particle radius $R$ and the inter-face distance $d$ between two particles. The lattice constant of the dimer chain is $l$. The orientation of the dimer is described by $\theta$, the angle between the long axis of the dimer and the chain direction. We use the finite-difference time-domain (FDTD) method to calculate the absorption cross section and the electric field intensity in the middle of the gap normalized to the incident field intensity $|E_g|^2$. The dielectric constant of Ag is obtained from the reference [48]. In the calculation, we have applied periodic boundary conditions along the infinite dimer chains and the perfect match layer conditions for the other boundaries for the dimer chain and all the boundaries for the single dimer. The incident light wavevector is perpendicular to the chain. The grids in the gaps are 0.5nm and the other grids are 1nm. The background index is 1 for the vacuum. Systems in other background show similar physical pictures.

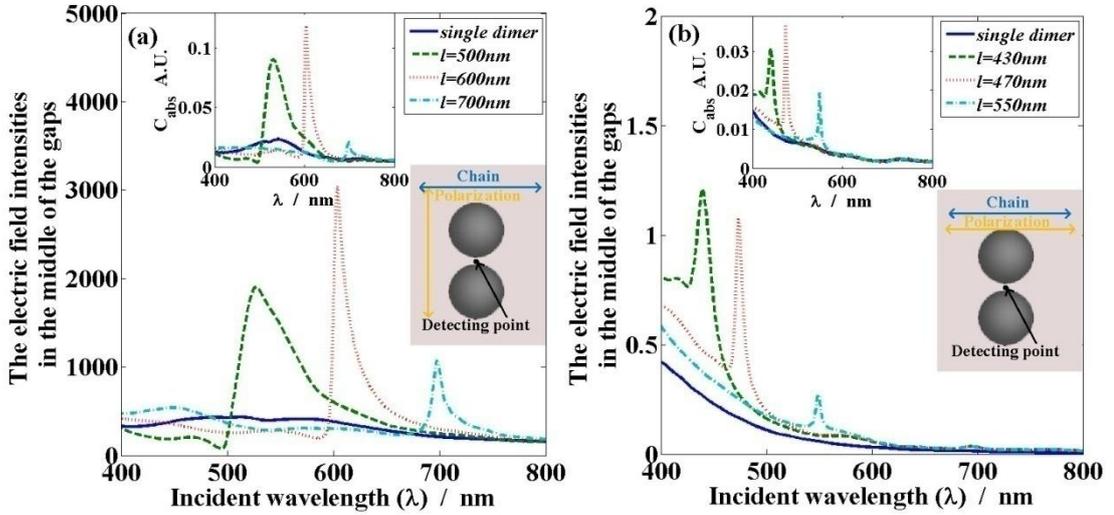

Figure 2. (Color online) The electric field intensities in the middle of the gaps $|E_g|^2$ vs. the incident wavelength of single dimer and three dimer chains. (a) The incident light polarizations are perpendicular to the chains and the lattice constants are 500nm, 600nm, 700nm, respectively; (b) The incident light polarizations are parallel to the chains and the lattice constants are 430nm, 470nm, 550nm, respectively; The absorption cross sections are shown in the inserts.

We first consider the cases of $\theta=90°$. The radius is $R=50$nm, and the inter-face distance is $d=10$nm. $|E_g|^2$ versus the incident wavelength of single dimer and three chains for the incident polarization perpendicular and parallel to the chain are shown in figures 2(a) and 2(b), respectively. The corresponding absorption cross sections are presented in the inserts in figures 2(a) and 2(b), respectively. All results (except those in figure 3) are plotted based on FDTD data.

From figure 2(a), we see the strong collective resonance and the width of the resonance becomes narrower with the lattice constant increasing from $l=500$nm to $l=700$nm. The absorption spectrum shows a similar lineshape as that for the gap electric field. The electric field intensity in the middle of the gap $|E_g|^2$ for each dimer chain is larger than the one for single dimer, which could be called collective hotspots. With increasing the lattice constants, $|E_g|^2$ first increases then decreases. The ratio between $|E_g|^2_{l=600nm,max}$ and $|E_g|^2_{single,max}$ could reach 6. The collective hotspots with very strong gap fields and an extremely narrow resonant width for $l$=600nm could be called anomalous collective hotspots, and the strong absorption with an extremely narrow resonant width for $l$=600nm could be called the anomalous absorption. To understand the physical mechanism of the anomalous collective hotspots, we use a generalized coupled dipole model to analysis the key features. The dimer chain may be viewed as two identical single nanoparticle chains (Chain A and Chain B) with a small inter-face distance $d$. We assume each metallic nanoparticle can be described by a dipole with the polarizability $\alpha$, the interactions among the nanoparticles lead to the following coupled equations:

$$P_{A_n} = \alpha[E_0 + \sum_{n'\neq n} G_k(x_n^A, x_{n'}^A)P_{A_{n'}} + \sum_{n'\neq n} G_k(x_n^A, x_{n'}^B)P_{B_{n'}} + \tilde{G}_k(x_n^A, x_n^B)P_{B_n}] \quad (1a)$$

$$P_{B_n} = \alpha[E_0 + \sum_{n'\neq n} G_k(x_n^B, x_{n'}^B)P_{B_{n'}} + \sum_{n'\neq n} G_k(x_n^B, x_{n'}^A)P_{A_{n'}} + \tilde{G}_k(x_n^B, x_n^A)P_{A_n}] \quad (1b)$$

where $E_0$ is the amplitude of the incident field,

$$\alpha = \frac{1-\left(\frac{1}{10}\right)(\varepsilon+\varepsilon_0)x^2}{\frac{4\pi}{V}\left(\frac{1}{3}+\frac{\varepsilon_0}{\varepsilon-\varepsilon_0}\right)-\frac{4\pi}{30V}(\varepsilon+10\varepsilon_0)x^2 - i\frac{2}{3}k^3}$$ [49], $\varepsilon$, $\varepsilon_0$ the dielectric constants of

bulk silver and vacuum, $k$ the wavevector of the incident light, $x=kR/2$, $V$ the volume of the nanoparticle, and $G_k(x_n^A, x_{n'}^A)$, $G_k(x_n^A, x_{n'}^B)$ describe the far field interaction between nanoparticles at a lattice position $nl$ of Chain A and nanoparticles at different lattice position $n'l$ of Chain A/B. $G_k(x_n^A, x_{n'}^B) \approx G_k(x_n^A, x_{n'}^A) = G_k(|x_n^A - x_{n'}^A|) \equiv G_k(ml)$, $m=n-n'$,

$$G_k(ml) = \left(\frac{k^2}{|ml|} + \frac{ik}{|ml|^2} - \frac{1}{|ml|^3}\right)e^{ik|ml|} \quad \text{or} \quad G_k(ml) = \left(-\frac{2ik}{|ml|^2} + \frac{2}{|ml|^3}\right)e^{ik|ml|} \quad \text{for the}$$

dipole orientation of the dimer perpendicular or parallel to the chain [50]. $\tilde{G}_k(x_n^A, x_n^B)$ describes the near field interaction between the nanoparticles at the same lattice position $nl$ of different Chain A and Chain B. Since the inter-face distance within a dimer is small and the condition for usual dipole interaction approximation (the inter-face distance larger than the radius of nanoparticles [19, 21, 50]) breaks down, more careful treatment is needed for $\tilde{G}_k(x_n^A, x_n^B)$, see appendix for details. After some calculation, in the case of large lattice constant and small inter-face distance within a dimer, we can obtain:

$$P_{A_n} + P_{B_n} = E_0/[1/2\alpha - S_k - \tilde{G}_k(d)/2] \equiv \alpha_{\text{eff}} E_0, \quad (2)$$

where $S_k=\Sigma_{m>0}2G_k(ml)$. The absorption spectra can be calculated as $C_{abs}=k\text{Im}[\alpha_{\text{eff}}]$. The optical absorption spectra based on the above equation shown in figure 3 agree quite well with those inserted in figure 2(a) based on FDTD simulations. The positions of dips in the curves inserted in figure 2(a) and figure 3 are the same, while the positions of the peaks in figure 3 have a slight shift within 2nm. Note that our generalized coupled dipole method works well even for the case of inter-face distance of 0.2R, under which the usual coupled dipole method breaks down. It is clearly seen that the absorption cross section and the electric field is determined by the properties of each nanoparticle (described by $\alpha$), the near field interaction (described by $\tilde{G}_k(d)$) and the far field interaction (described by $S_k$ or $G_k(x_n^A, x_{n'}^A)$, $G_k(x_n^A, x_{n'}^B)$).

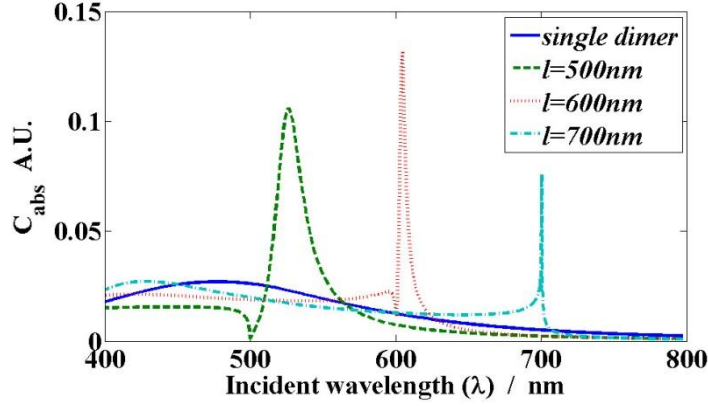

Figure 3. (Color online) The absorption spectra of a single dimer and dimer chains with lattice constant $l$=500nm, 600nm, 700nm. The inter-face distance in the dimer is $d$=10nm.

From Eq. (2), one sees that the resonant wavelength is determined by the equation $\text{Re}[1/2\alpha - S_k - \tilde{G}_k(d)/2] = 0$, which is very close to the resonant wavelength in figure 2(a) based on the FDTD simulation. Moreover, the partial cancellation of the imaginary part of the inverse of the nanoparticle polarizability $1/\alpha$ and those of $2S_k$ and $\tilde{G}_k(d)$ leads to the extremely narrow resonance width. There is also a lattice constant for maximal absorption [19] and the gap field enhancement. From Eq. (2), we see that each dimer composed of two particles with the inter-face distance $d$ could be treated as a unit with dipole moment $P_A+P_B$ (which could be described by the effective polarizability $\alpha_D(d)$ defined by $1/\alpha_D = [1/2\alpha - \tilde{G}_k(d)/2]$). Firstly, the volume doubling from $V$ (for a particle) to $2V$ (for a dimer) leads to the change from $1/\alpha$ to $1/2\alpha$, and $\tilde{G}_k(d)$ takes the near field interaction into account. Further lattice/collective effect is included in $S_k$.

The strong gap field in the dimer chain is the dual effects of inter-particle interaction within the dimer and the far field interaction from other dimers. From figure 2(a), we see that the gap field enhancement factor of single dimer is about 500. Basically, the collective effect from other dimers leads to a large effective field on the dimer. The effective local field on the dimer $E_D$ is the sum of the incident field and the field from other units (with effective dipoles), i.e.,

$$E_D = E_0 + \frac{S_k}{1/\alpha_D - S_k} E_0. \qquad (3)$$

Important physical pictures can be drawn from above equation. It is seen that the interference between the incident field (the background channel) and the induced internal field (the resonant channel) may lead to the Fano-type absorption lineshape [51, 52]. In particular, at Wood anomaly [53, 54] with $l=\lambda$, $S_k \approx -\ln[4(1-\cos(x'))^2]/2x' \to \infty$ (as $x'=kl \to 2\pi$) [55], which leads to the destruction interference and the complete cancellation between the incident field and induced internal field on the dimer. In fact, the local field intensity on the dimer $|E_D|^2$ can be written in the following form near Wood anomaly:

$$|E_D|^2 = \left|\frac{1}{1-\alpha_D S_k}E_0\right|^2 \approx \frac{(A^2+B^2)\pi^2 z^2}{(1+A\pi z + B\beta\pi^2 z^2)^2 + B^2\pi^2 z^2}|E_0|^2, \quad (4)$$

with the variables/parameters defined by $\frac{1}{\alpha_D} = A+iB$, $S_k = -\frac{1}{\pi z} - i\beta$,

$z = \frac{1}{\ln|2\pi(l/\lambda - 1)|}$. The above equation can be written in the Fano function [56] as

$$|E_D|^2 \propto \frac{(\varepsilon+q)^2}{\varepsilon^2+1}, \quad \text{with} \quad \varepsilon = (Kz+A)/Q, \quad q = -\frac{A}{Q}, \quad K = (A^2+B^2+2B\beta)\pi,$$

$Q = \sqrt{B^2+2B\beta}$. The absorption can also be written in the same form $C_{abs} \propto \frac{(\varepsilon+q)^2}{\varepsilon^2+1}$. It is seen that the Fano factor depends on both the local structure (i.e. $\alpha_D$) and the long rang interaction (i.e., $S_k$). The Fano zero due to the destructive interference is related to Wood anomaly. The Fano resonance appears in the condition of $\text{Re}[\alpha_D S_k]=1$, showing the combination effects of the near field interaction and the collective behavior. The above perturbative results are valid in a wider range and clear Fano lineshapes can be seen in the curves of the absorption and the gap field versus $z$ in figure 4. The dimer chains with a smaller $d$ (6nm) have smaller Fano factor $q$, showing more pronounced Fano-type effect (the asymmetric lineshape). Figure 4 shows very good fitting of FDTD data based on Fano function. The gap field in the dimer unit can be viewed as excited by the new effective field $E_D$, which is larger than the incident field and results in further enhancement of the gap field with enhancement factor 3000.

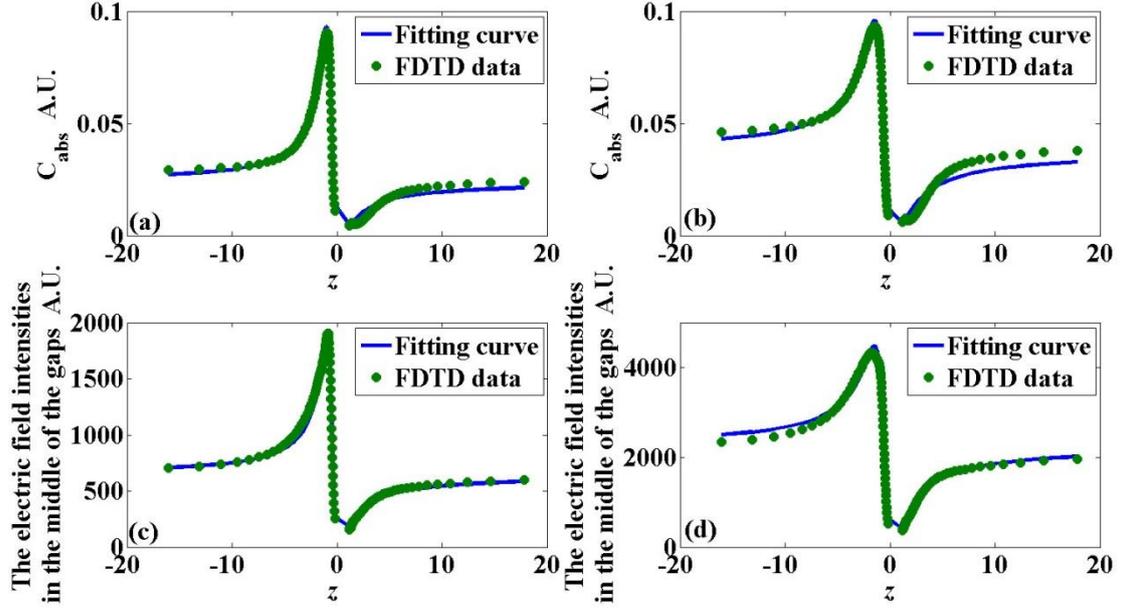

Figure 4. (Color online) The absorption spectra (a) (b) and gap fields (c) (d) vs. $z$ for of 1D dimer chain with lattice constant $l$=500nm. The inter-face distance in the dimer is $d$=10nm (for (a) and (c)) and $d$=6nm (for (b) and (d)). The solid lines are the results from fitting based on the Fano function and the dots represent the results from FDTD calculations.

Thus, we obtain the main physical picture: a dimer unit in the chain may be viewed as a single dimer in the presence of the effective field from the collective action of other dimers. Therefore the lineshapes for the field enhancement and the absorption may look similar. Compared with the hotspot of single dimer, the collective hotspots appear at a red shifted wavelength and the gap field is stronger. Moreover, for specific lattice constant, anomalous collective hotspots with extremely narrow resonance width and a giant gap field can be obtained. Our generalized coupled dipole method, which incorporates both the near field and the far field interaction, not only captures the main physical picture but also clarifies the relation between Wood anomaly and Fano resonance as shown in figure 4. (*Note that Wood anomaly leads to a singularity, while the Fano function has no singularity.*)

We next consider the case of the incident light polarization along the chain. From above discussion we see that the anomalous absorption/hotspots, which is due to the partial cancellation between the imaginary of $1/\alpha_D$ and $S_k$, depends on the lattice constant, the incident field polarization, and the dimer orientation. Therefore, here we use different sets of lattice constants to see the anomalous absorption/hotspots. As shown from figure 2(b) with $l$=430nm, 470nm and 550nm, the collective action also makes the lineshape of the absorption narrower. This behavior is quite different from the usual 1D single particle chain. It is the inter-particle interaction within a dimer $\tilde{G}_k(d)$ that leads to a different cancellation mechanism. Since the incident light polarization is parallel to the chain, the convergence of $S_k$ results in the disappearance of the complete destructive interference. Also the field enhancement due to strong local effect would not appear.

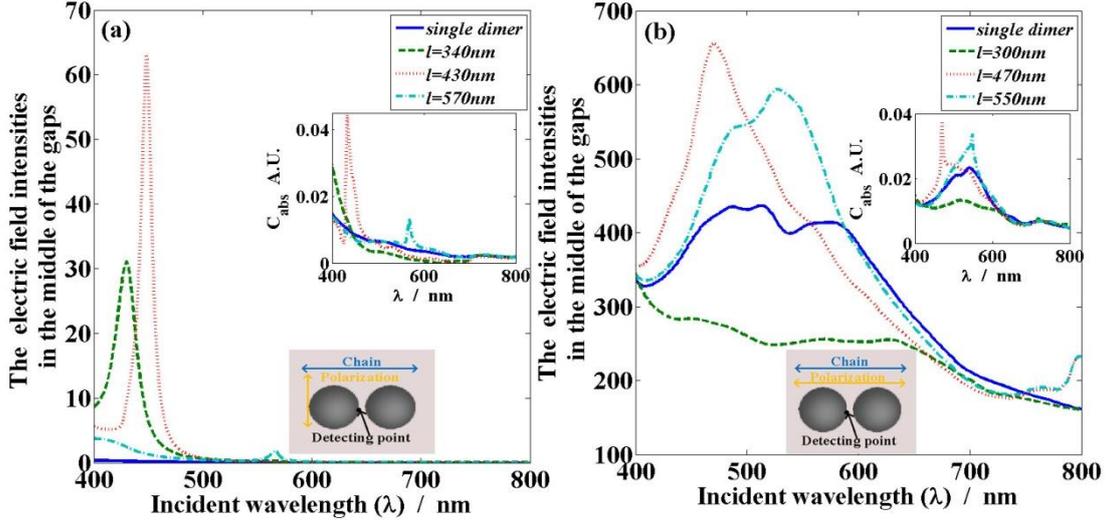

Figure 5. (Color online) The electric field intensities in the middle of the gaps $|E_g|^2$ vs. the incident wavelength of single dimer and three chains. (a) The incident field polarization is perpendicular to the chain and the lattice constants are 340nm, 430nm, 570nm, respectively; (b) The incident field polarization is parallel to the chain and the lattice constants are 300nm, 470nm, 550nm, respectively; The absorption cross sections are shown in the inserts.

As we have discussed, the combination effects of the near field and the far field interactions play the essential roles in the optical properties. Therefore, the relative orientation of the dimer with respect to the chain direction is an important factor. We then consider the case of the long axis of the dimer parallel to the chain $\theta=0°$. The results for the incident light polarization perpendicular to the chain with lattice constant $l$=340nm, 430nm and 570nm are shown in figure 5(a). In this case of the incident light polarization perpendicular to the long axis of dimers, anomalous collective hotspots mainly originate from the collective action of the chains and the enhancement would not be as large as the one in the structure with the polarization parallel to the long axis of the dimer. It is the lack of the strong near field effect that leads to weaker field enhancement.

From figure 5(b), we see that $|E_g|^2$ for chains with lattice constants in the regime from 300nm to 570nm is around the value of $|E_g|^2$ for single dimer. It is consistent with the picture that the field enhancement is mainly attributed to the near field effect (the polarization is along the gap direction) and the collective effect plays less important role for the case of the incident polarization parallel to the chain direction. Also the imaginary part of $S_k$ for $l$=300nm is very different from those for $l$=470nm, 550nm, which leads to the difference in the profiles.

We next give more comparison of different dimer orientations with angles between the long axis of the dimer and the chain direction $\theta$=0°, 30°, 45°, 60°, 90°. Here we only consider the incident light polarized along the long axis of the dimers, since the gap field is strong for these cases. For the dimer chain with different angle $\theta$, the local structure is the same (with the incident polarization parallel to the long axis of the dimer) and the global structure is the same (with the same period), yet different correlation between the local structure and global structure, i.e., $\theta$, leads to different physical consequence. It is interesting to see from figure 6 that the gap field for the $\theta$=60° is strongest, while the absorption is strongest for $\theta$=90°. For the absorption, the dimer may be viewed as one object described by an effective polarizability

$\alpha_D$ based on our generalized coupled dipole model. As $\theta$ changing from 0° to 90°, the absorption increases due to the decrease of Im[$1/\alpha_D-S_k$]. While for the enhanced gap field, the local structure of the dimer is relevant and additional correction beyond the simple effective description is needed. In this situation, the local field (due to other dimers) component perpendicular to the long axis of the dimer ($\propto \sin(\theta)\cos(\theta)$) leads to another contribution to the gap field enhancement. Then an optimal angle $\theta$ for the strongest gap field enhancement may appear.

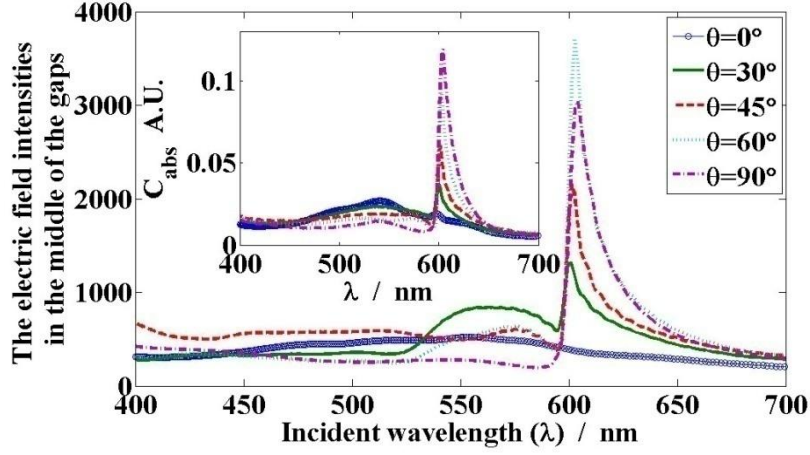

Figure 6. (Color online) The comparison of absorption and gap fields for different dimer orientations with angles between the long axis of the dimer and the chain direction $\theta$=0°, 30°, 45°, 60°, 90°. The incident light is polarized along the long axes of the dimers.

The inter-face distance $d$ affects the near field interaction and also plays an important role in the optical response. Figure 7 shows the absorption spectra and $|E_g|^2$ versus the incident wavelength of 1D dimer chain with different inter-face distances of 2nm, 6nm and 10nm. The lattice constant is fixed at 600nm. The polarization as well as the long axis of the dimer is perpendicular to the chain. It is seen from figure 7(a) that the lineshape of the absorption becomes broader and the peak has a red-shift with the decrease of $d$. In fact, it was shown in [55] that the resonant wavelength of 1D nanosphere chain is $\lambda_{res} = \lambda_{Wood} + \Delta\lambda$, with $\Delta\lambda \propto e^{-C\text{Re}[1/\alpha(\lambda=l)]}$, $\alpha$ the polarizability of a nanosphere. In our case, the interaction between the nanoparticle within a dimer leads to the inter-face distance dependent $\alpha_D(d)$ and $\Delta\lambda \propto e^{-C\text{Re}[1/\alpha(\lambda=l)]} \cdot e^{C\text{Re}[\tilde{G}_{k=2\pi/l}(d)]/2}$. In general, with the change of $d$, the $\alpha_D(d)$ ($\tilde{G}(d)$) changes, which makes the absorption spectra broader and the peak red-shifted. It is interesting to see that the peak value of the absorption has little change with the change of $d$, which is quite different from the single dimer behavior. The near field interaction in a single dimer leads to the increase of peak values of the absorption with the decrease of $d$ [57]. But in the 1D dimer chain, the change of $\alpha_D(d)$ results in the changing of the effective field on the dimer. These two effects makes lilttle change of peak values of the absorption. The gap field intensity $|E_g|^2$ increases sharply with $d$ reduced as shown in figure 7(b). $|E_g|^2_{d=2nm,max}$ is about 70000 times the value of the incident field intensity, which is about 9 times the value of the one for $d$=6nm, and 23 times the value of the one for $d$=10nm. The gap field in the collective hotspots is much stronger than that in a single dimer. The gap field enhancement in the dimer chain with $l$=600nm is about 3000/8000/70000, which is 6 times/4 times/2.5 times as that in a single dimer for the gap of size $d$=10nm/6nm/2nm. Further improvement of the gap field enhancement of small gap sizes would be achieved by choosing an appropriate lattice constant.

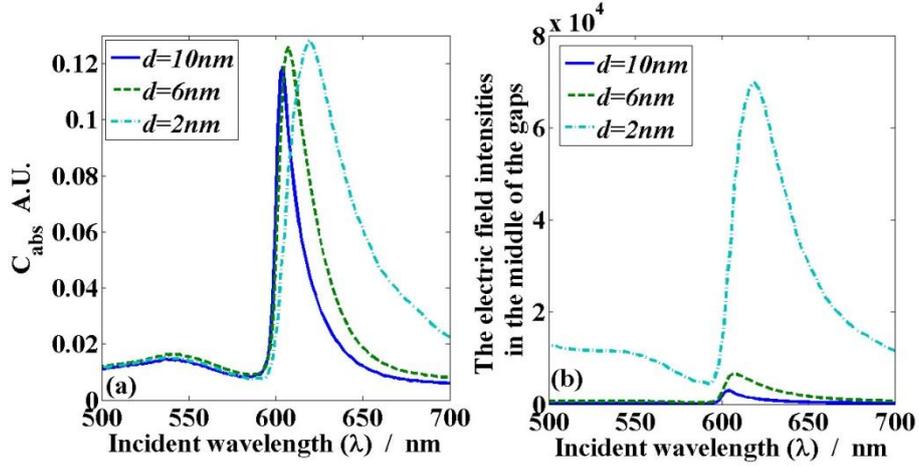

Figure 7. (Color online) The absorption spectra (a) and $|E_g|^2$ vs. the incident wavelength (b) of dimer chains with inter-face distances in the dimer $d$=2nm, 6nm, 10nm, respectively. The lattice constant $l$=600nm.

## 3. Conclusions

We have found that the interplay between LSPs induced strong near field and the collective plasmon resonance in the Ag nanoparticle dimer chains leads to Fano-type resonance and anomalous collective hotspots with large gap fields and an extremely narrow resonance width. The incident field polarization, the lattice constant, the dimer structure, and the correlation between the local structure and the global structure affect the optical response, leading to anomalous absorption and much stronger gap field than that in a single dimer. For the gap size of 10nm in each dimer, the gap field intensity enhancement in the collective hotspots could reach to 3000 (3600) times, which is about 6 (7) times as large as that of single dimer for $\theta$=90° ($\theta$=60°). The tunability of the lattice, dimer structures provides us more opportunities to manipulate the collective hotspots (with desired resonant frequency and strong local field), which have many applications in SERS and biosensors.

**Acknowledgement:**

This work was partially supported by the National Natural Science Foundations of China (Nos. 11174042 and 11374039), the National Basic Research Program of China (973 Program) under Grant Nos. 2011CB922204, 2013CB632805, and the China Postdoctoral Science Foundation under Grant No. 2014M560919.

**Appendix: Calculation of near field interaction between nanoparticles within a dimer**

The nanoparticle A (B) in the dimer at position $\vec{r}_A = (x_n, 0, R+d/2)$

($\vec{r}_B = (x_n, 0, R-d/2)$) is composed a series of disks with width $dz$ at position $z$ for A ($z'$ for B). In our parameter regime, we may assume the dipole moment of each disk of nanoparticle A (B) to be $d\vec{p}_A(z) = pS(z)dz\vec{z}$ ($d\vec{p}_B(z') = pS(z')dz'\vec{z}$), $p$ the density of dipole moment, $S(z)$($S(z')$) the area of the disk at position $z$ ($z'$), $\vec{z}$ the unit vector along $z$ axis (the long axis of the dimer). The total dipole moment of nanoparticle A (B) is $\vec{P}_A = \int d\vec{p}_A$ ($\vec{P}_B = \int d\vec{p}_B$). The interaction between nanoparticles A and B can be calculated as

$$\tilde{G}_k(x_n^A, x_n^B) = \tilde{G}_k(d) = \iint d\vec{p}_A(z) \cdot \{[3\vec{z}(\vec{z} \cdot d\vec{p}_B) - d\vec{p}_B](\frac{1}{r^3} - \frac{ik}{r^2})e^{ikr}\}/(|\vec{P}_A||\vec{P}_B|) \quad , \quad \text{with}$$

$r = |z - z'|$.